\documentclass[11pt,letterpaper]{mystyle}

\usepackage[all]{hypcap}

\PassOptionsToPackage{comma,numbers,compress}{natbib}

\usepackage[comma,authoryear,compress]{natbib}
\usepackage{hyperref}[citecolor=lightblue]

\hypersetup{
    colorlinks = true,
    citecolor = {YaleBlue},
}

\usepackage{algorithm}
\usepackage{algorithmicx}
\usepackage{algpseudocode}
\usepackage{microtype}
\usepackage{graphicx}
\expandafter\def\csname ver@subfig.sty\endcsname{}
\usepackage{booktabs} %
\usepackage{float}
\usepackage{bigstrut}

\usepackage{amsmath}
\usepackage{amssymb}
\usepackage{mathtools}
\usepackage{amsthm}
\usepackage{mathrsfs}
\usepackage{nicefrac}
\usepackage{dsfont}
\usepackage{enumitem}
\usepackage{subcaption}
\usepackage{graphicx,subfig}
\usepackage{cleveref}
\usepackage{bxcoloremoji}

\usepackage{float}

\usepackage[utf8]{inputenc} %
\usepackage[T1]{fontenc}    %
\usepackage{hyperref}       %
\usepackage{url}            %
\usepackage{booktabs}       %
\usepackage{amsfonts}       %
\usepackage{nicefrac}       %
\usepackage{microtype}      %
\usepackage{xcolor}         %
\usepackage{graphicx}
\usepackage{subcaption}
\usepackage{amssymb}
\usepackage{fdsymbol}
\usepackage{wrapfig}
\usepackage{lipsum}
\usepackage{enumitem}
\usepackage{stackengine}
\usepackage[font=small,labelfont=bf]{caption}
\usepackage{color}
\usepackage{adjustbox}

\newcommand{\x}{\mathbf{x}}

\definecolor{blanchedalmond}{rgb}{1.0, 0.92, 0.8}
\definecolor{carmine}{rgb}{0.59, 0.0, 0.09}
\definecolor{lightblue}{rgb}{0.22,0.45,0.70}%

\renewcommand{\mathbf}{\boldsymbol}

\makeatletter
\def\Ddots{\mathinner{\mkern1mu\raise\p@
\vbox{\kern7\p@\hbox{.}}\mkern2mu
\raise4\p@\hbox{.}\mkern2mu\raise7\p@\hbox{.}\mkern1mu}}
\makeatother

\definecolor{amaranth}{rgb}{0.9, 0.17, 0.31}
\definecolor{antiquebrass}{rgb}{0.8, 0.58, 0.46}
\definecolor{antiquefuchsia}{rgb}{0.57, 0.36, 0.51}
\definecolor{chromeyellow}{rgb}{0.31, 0.47, 0.26}

\newcommand{\github}{\raisebox{-1.5pt}{\includegraphics[height=1.05em]{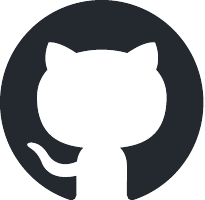}}}

\newtcolorbox{AIbox}[2][]{aibox,title=#2,#1}
\definecolor{lightblue}{rgb}{0.22,0.45,0.70}%
\definecolor{Gray}{gray}{0.95}
\definecolor{Cornsilk}{rgb}{1.0, 0.97, 0.86}

\usepackage{amsmath}

\usepackage[all]{hypcap}

\title{\resizebox{1.55in}{\height}{\textbf{METAGENE-1}}: Metagenomic Foundation Model\\for Pandemic Monitoring}

\runningtitle{\textbf{METAGENE-1}: Metagenomic Foundation Model for Pandemic Monitoring}

\author[1]{Ollie Liu}
\author[2]{Sami Jaghouar}
\author[2]{Johannes Hagemann}
\author[1]{Shangshang Wang}
\author[1]{\\Jason Wiemels}
\author[3]{Jeff Kaufman}
\author[1]{Willie Neiswanger}

\affil[1]{University of Southern California}
\affil[2]{Prime Intellect}
\affil[3]{Nucleic Acid Observatory}

\correspondingauthor{Ollie Liu \href{mailto:me@ollieliu.com}{me@ollieliu.com}; Willie Neiswanger \href{mailto:neiswang@usc.edu}{neiswang@usc.edu}}

\begin{abstract}
We pretrain METAGENE-1, a 7-billion-parameter autoregressive transformer model, which we refer to as a \textit{metagenomic foundation model}, on a novel corpus of diverse metagenomic DNA and RNA sequences comprising over 1.5 trillion base pairs. This dataset is sourced from a large collection of human wastewater samples, processed and sequenced using deep metagenomic (next-generation) sequencing methods. Unlike genomic models that focus on individual genomes or curated sets of specific species, the aim of METAGENE-1 is to capture the full distribution of genomic information present within this wastewater, to aid in tasks relevant to pandemic monitoring and pathogen detection. We carry out byte-pair encoding (BPE) tokenization on our dataset, tailored for metagenomic sequences, and then pretrain our model. In this paper, we first detail the pretraining dataset, tokenization strategy, and model architecture, highlighting the considerations and design choices that enable the effective modeling of metagenomic data. We then show results of pretraining this model on our metagenomic dataset, providing details about our losses, system metrics, and training stability over the course of pretraining. Finally, we demonstrate the performance of METAGENE-1, which achieves state-of-the-art results on a set of genomic benchmarks and new evaluations focused on human-pathogen detection and genomic sequence embedding, showcasing its potential for public health applications in pandemic monitoring, biosurveillance, and early detection of emerging health threats.
\vspace{5mm}

\coloremojicode{1F310} \textbf{Website}: \href{https://metagene.ai}{\textbf{metagene.ai}}

\coloremojicode{1F917} \textbf{Model Weights}: \href{https://huggingface.co/metagene-ai}{\textbf{huggingface.co/metagene-ai}}

\github{} \textbf{Code Repository}: \href{https://github.com/metagene-ai}{\textbf{github.com/metagene-ai}}
\end{abstract}

\begin{document}
\maketitle
\vspace{3mm}
\vspace{-4mm}
\section{Introduction}
\label{sec:intro}
\vspace{-1mm}

The development of large language models trained on internet-scale text datasets has revolutionized natural language processing, finding increasingly broad applications across numerous domains.
In recent years, this modeling technology has been adapted to genomic sequences---\textit{e.g.}, DNA or RNA strands that carry genetic information---leveraging the wealth of data generated by advances in genome sequencing over the past few decades~\citep{ji2021dnabert,nguyen2024hyenadna,dalla2023nucleotide,zhou2023dnabert,fishman2023gena}.
These large genomic models aim to harness modeling power for tasks such as genome classification, phenotype prediction, gene network inference, human genome analysis, and biological design for medical and therapeutic applications.
To date, most of these models have been trained on human genomes or on curated collections of genomes from selected species~\citep{consens2023transformers, benegas2024genomic}.

Parallel to these developments, there has been significant work on large-scale health monitoring driven largely by widespread public health crises, such as the COVID-19 pandemic~\citep{salomon2021us, reinhart2021open}.
One notable example of this is the genomic monitoring of \emph{wastewater}, which involves sequencing material from samples of municipal sewage~\citep{farkas2020wastewater, consortium2021global}.
Wastewater contains a complex mix of organic materials generated from human activities and, when collected across multiple time points and locations, can reveal valuable information about the microbiome at a societal scale~\citep{bogler2020rethinking,levy2023wastewater}.
Consequently, there have been various efforts to collect wastewater and sequence \emph{metagenomic information}, i.e., information about the diverse collections of organisms and organic material present in these samples~\citep{medema2020implementation, mao2020potential,mcclary2021sars}.
A key motivation for much of this work is the potential to track the prevalence of human pathogens, effectively creating an early warning system for pandemics.
Multiple ongoing initiatives are collecting vast amounts of metagenomic information to monitor genomic trends, estimate the prevalence of sequences of interest, and detect new or emerging potential pathogens~\citep{consortium2021global,keshaviah2021developing,levy2023wastewater}.

These wastewater metagenomic sequencing efforts present two significant opportunities.
First, they provide a novel and rich source of metagenomic data, rivaling the scale of datasets used to pretrain large language models (\textit{i.e.}, trillions of nucleic acid base pairs), encompassing highly diverse genomic information across the broad human-adjacent microbiome~\citep{breitwieser2019review,tisza2021catalog}.
This metagenomic data often exhibits unique distributional characteristics in terms of genomic sequence length, heterogeneity, and composition/type of organisms, distinguishing it from previous genome modeling datasets.
Second, this data opens up a new domain area for downstream applications of foundation models trained on this information.
Such models could be fine-tuned for various tasks crucial to pathogen monitoring, including tracking frequencies, trends, and growth of different sequence types; representation learning and embedding for sequenced metagenomic reads; sequence alignment, error-correction, and infilling; and human pathogen detection and taxonomic classification~\citep{consortium2021global}.

\begin{figure}[t]
    \centering
    \includegraphics[width=\textwidth]{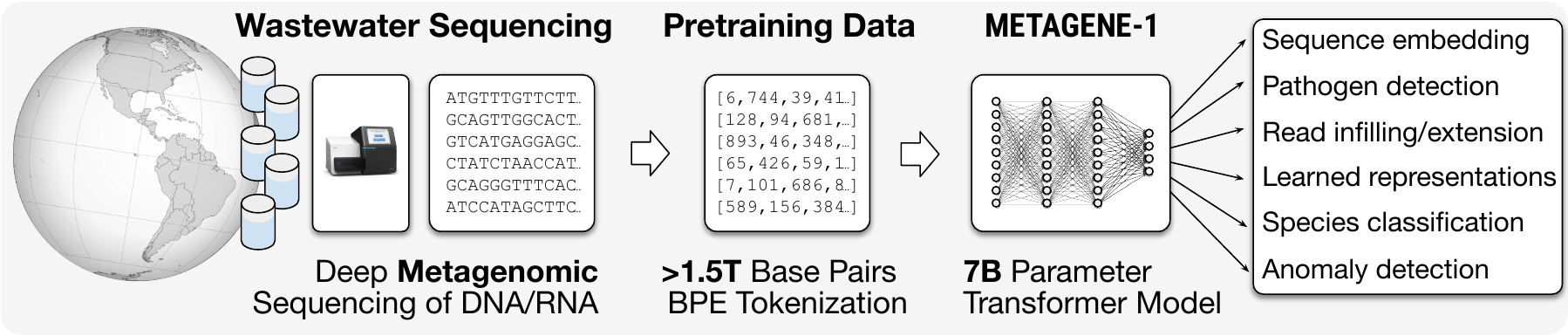}
    \caption{
    \textbf{Overview of METAGENE-1 and applications}. Wastewater samples are collected and undergo deep metagenomic sequencing to generate DNA and RNA sequences totaling over 1.5 trillion base pairs. These sequences are tokenized using byte-pair encoding (BPE) to create the pretraining dataset. The data is used to train METAGENE-1, a 7B-parameter transformer model that enables a wide range of metagenomic analysis and monitoring applications.}
    \label{fig:overview-figure}
    \vspace{-1mm}
\end{figure}

In this paper, we take an initial step toward developing a metagenomic foundation model by pretraining a model on a large, new dataset sequenced from \textit{wastewater}.
This metagenomic dataset, which has never before been used for model training, provides a unique resource for modeling the broad distribution of sequences present in the human microbiome.
Specifically, we pretrain a 7-billion-parameter autoregressive transformer model, which we refer to as METAGENE-1, on a diverse corpus of DNA and RNA sequences comprising over 1.5 trillion base pairs sourced from wastewater samples, which were processed and sequenced using deep metagenomic (next-generation) sequencing~\citep{bragg2014metagenomics, consortium2021global}.
This dataset, comprising short uncurated sequences from tens of thousands of species, allows METAGENE-1 to excel at representing the complexities of microbial and viral diversity, providing unique advantages in biosurveillance applications.
METAGENE-1 adopts a decoder-style language model architecture, similar to those found in the GPT and Llama families of models~\citep{radford2019language,llama2}, which we describe and motivate in more detail in Sec.~\ref{sec:metagene-architecture}.
This choice allows us to take advantage of the broad (and rapidly growing) ecosystem of techniques and infrastructure focused on this class of models. An overview of METAGENE-1 data, model architecture, and applications is shown in Figure~\ref{fig:overview-figure}.

In the following sections, we first describe our metagenomic dataset and detail the tokenization strategy used to process the sequence data.
We then provide comprehensive details of the METAGENE-1 model architecture and of the pretraining process on our dataset.
Subsequently, we develop, and demonstrate our model's performance, on pathogen detection and metagenomic embedding benchmarks.
METAGENE-1 achieves state-of-the-art performance on these and other standard genomic evaluation tasks---designed to evaluate models trained on human and animal genomes---highlighting its generalization capabilities.
As an initial demonstration of the downstream application potential, we construct an anomaly detection scenario, and show that METAGENE-1 performs well on this out-of-distribution detection task.
We hope our paper serves as an initial step toward a foundation model for metagenomic data, which in the future can be fine-tuned to aid in public health applications such as pathogen monitoring and early detection of emerging health threats.

\section{Related Work}
\label{sec:related_work}

Language models trained on genomic sequences have been an area of active research, with many aiming to train on long DNA sequences from specific species, gained from publicly available sources.
For instance, models such as DNABERT~\citep{ji2021dnabert}, HyenaDNA~\citep{nguyen2024hyenadna}, GROVER~\citep{sanabria2024dna}, and Caduceus~\citep{schiff2024caduceus} are examples primarily trained on long sequences of \textit{human DNA}. These models typically use encoder-based architectures or decoder-only non-transformer architectures, aiming to handle long sequence lengths. For tokenization, these initial human-focused genome models have commonly employed either $k$-mer tokenization (with fixed values like $k$=3) or single-nucleobase tokenization.

Recently, the scope of genomic models has expanded to include multi-species datasets, with models like DNABERT-2~\citep{zhou2023dnabert}, NucleotideTransformer~\citep{dalla2023nucleotide}, GENA-LM~\citep{fishman2023gena}, SpliceBERT~\citep{chen2023self}, and DNAGPT~\citep{zhang2023dnagpt} being trained on a mix of human genome data and manually curated sets from other species (for example, mixes of species from a taxonomic class, such as collections of mammals). Some of these models have also explored alternative tokenization strategies, such as byte-pair encoding, learned for their particular genomic distributions~\citep{zhou2023dnabert, fishman2023gena, sanabria2024dna, zhou2024dnabert}.

Our metagenomic foundation model differs from these prior works in a few important ways. First, our pretraining dataset comprises shorter metagenomic sequences (arising from metagenomic next-generation/massively-parallel sequencing methods) performed on samples of human wastewater collected across many locations; these samples contain potentially tens-of-thousands of species across a wide range of taxonomic ranks, and capture a representative distribution of the full human-adjacent microbiome. This includes both recognized species and many unknown or unclassified sequences (see Sec.~\ref{sec:dataset}). Another distinction is the model architecture: we use a decoder-only transformer model, akin to the Llama and GPT model families, which we further motivate in Sec.~\ref{sec:metagene-architecture}.

\section{METAGENE-1: Metagenomic Foundation Model}
\label{sec:metagene-model}

We pretrain a 7-billion-parameter autoregressive transformer language model, referred to as METAGENE-1, on a novel corpus of diverse metagenomic DNA and RNA sequences comprising over 1.5 trillion base pairs.
This dataset is sourced from a diverse set of human wastewater samples, which were processed and sequenced using deep metagenomic (next-generation) sequencing methods.
Before training, we carry out byte-pair encoding (BPE) tokenization on our dataset, tailored for these nucleic acid sequences.
The following sections provide detailed descriptions of the pretraining dataset, tokenization strategy, and model architecture, highlighting the considerations and design choices that enable the effective modeling of metagenomic data.

\subsection{Metagenomic Dataset}
\label{sec:dataset}

One of the goals of our metagenomic foundation model is to train on a genomic dataset that captures the immense diversity of the microbiome surrounding humans.
To achieve this, we leverage a newly collected metagenomic dataset---never before used in model training---comprising material from a broad range of organisms, including bacteria, viruses, cells from human and other eukaryotes, and a diverse array of other species, which was collected via \textit{metagenomic sequencing of human wastewater} (i.e., municipal influent).
This approach contrasts with prior genomic sequence models, which often focus on curated collections of specific (known) species or genomic types.
By incorporating DNA and RNA sequences collected from wastewater, we aim to model the complexity of microbial and viral interactions in human-associated environments.

\begin{figure}[t]
    \centering
    \includegraphics[width=\textwidth]{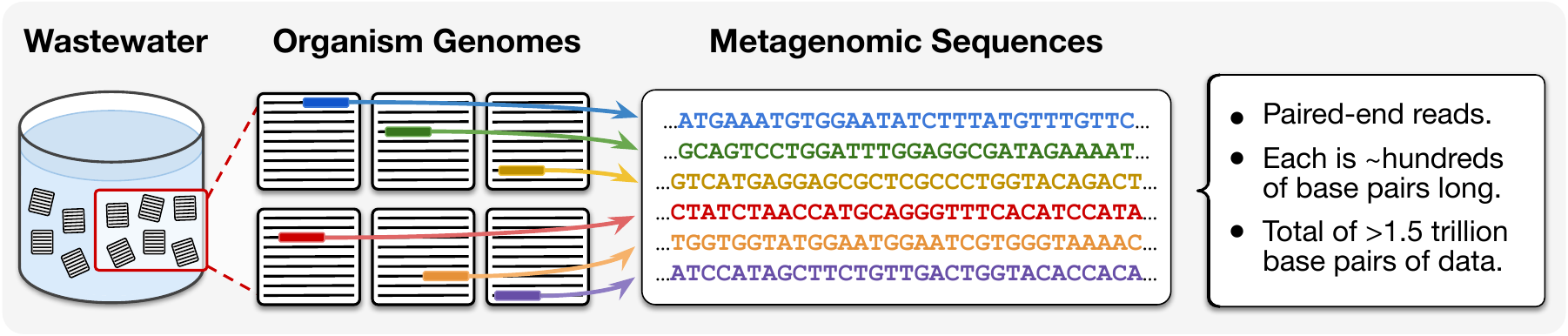}
    \caption{\textbf{Overview of the metagenomic data collection and sequencing pipeline for model pretraining}.
    The process begins with the collection of wastewater (left), which contains genomic fragments from a diverse collection (\textit{e.g.}, tens of thousands) of constituent organisms (center). These samples are processed via high-throughput metagenomic sequencing to produce millions of paired-end reads (right), each consisting of hundreds of base pairs. The complete dataset comprises over 1.5 trillion base pairs of metagenomic sequences used for model pretraining.}
    \label{fig:metagenomic-data}
    \vspace{-1mm}
\end{figure}

The dataset was generated using deep metagenomic sequencing, specifically leveraging Illumina sequencing technology, commonly referred to as next-generation sequencing (NGS) or high-throughput sequencing, in which billions of nucleic acid fragments are simultaneously sequenced in a massively parallel manner.
This method produces paired-end reads, where each read consists of two contiguous sequences of base pairs from opposite ends of a DNA or RNA fragment\footnote{Where RNA sequences are first converted into DNA via reverse transcription. $^2$\url{https://bondlsc.missouri.edu/person/marc-johnson}. $^3$\url{https://jasonrothman.weebly.com/}}.
Paired-end reads can offer advantages in accuracy and alignment over single-end reads, particularly for complex metagenomic samples. Notably, the nature of metagenomic NGS results in much shorter reads compared to datasets used in many previous large genomic models. In our dataset, most reads range from 100 to 300 base pairs in length (after adapter removal and quality trimming), which introduces unique challenges for modeling, but also provides a rich diversity and large set of biological information. We illustrate this metagenomic data collection and sequencing pipeline in Figure~\ref{fig:metagenomic-data}.

This metagenomic sequence corpus was collected over a six-month period by the Nucleic Acid Observatory (NAO)~\citep{consortium2021global} in collaboration with partners (Marc Johnson and Clayton Rushford at the University of Missouri\footnotemark{} and Jason Rothman in Katrine Whiteson’s lab\footnotemark{} at the University of California, Irvine).
Samples of wastewater were sourced from multiple locations across the United States, in particular from cities in California and Missouri.
After wastewater samples were collected, the material was filtered and nucleic acids extracted~\citep{rothman2021rna,robinson2022defining} before undergoing metagenomic sequencing.
In full, the metagenomic dataset for pretraining comprises over 1.5 trillion base pairs.
Our hope is that this careful sampling and processing approach yields a clean dataset for sequence modeling, which captures a wide array of genomic content, offering a strong foundation for the training of METAGENE-1.

\begin{figure}[h]
    \centering
    \vspace{-10mm}
    \includegraphics[width=0.7\textwidth]{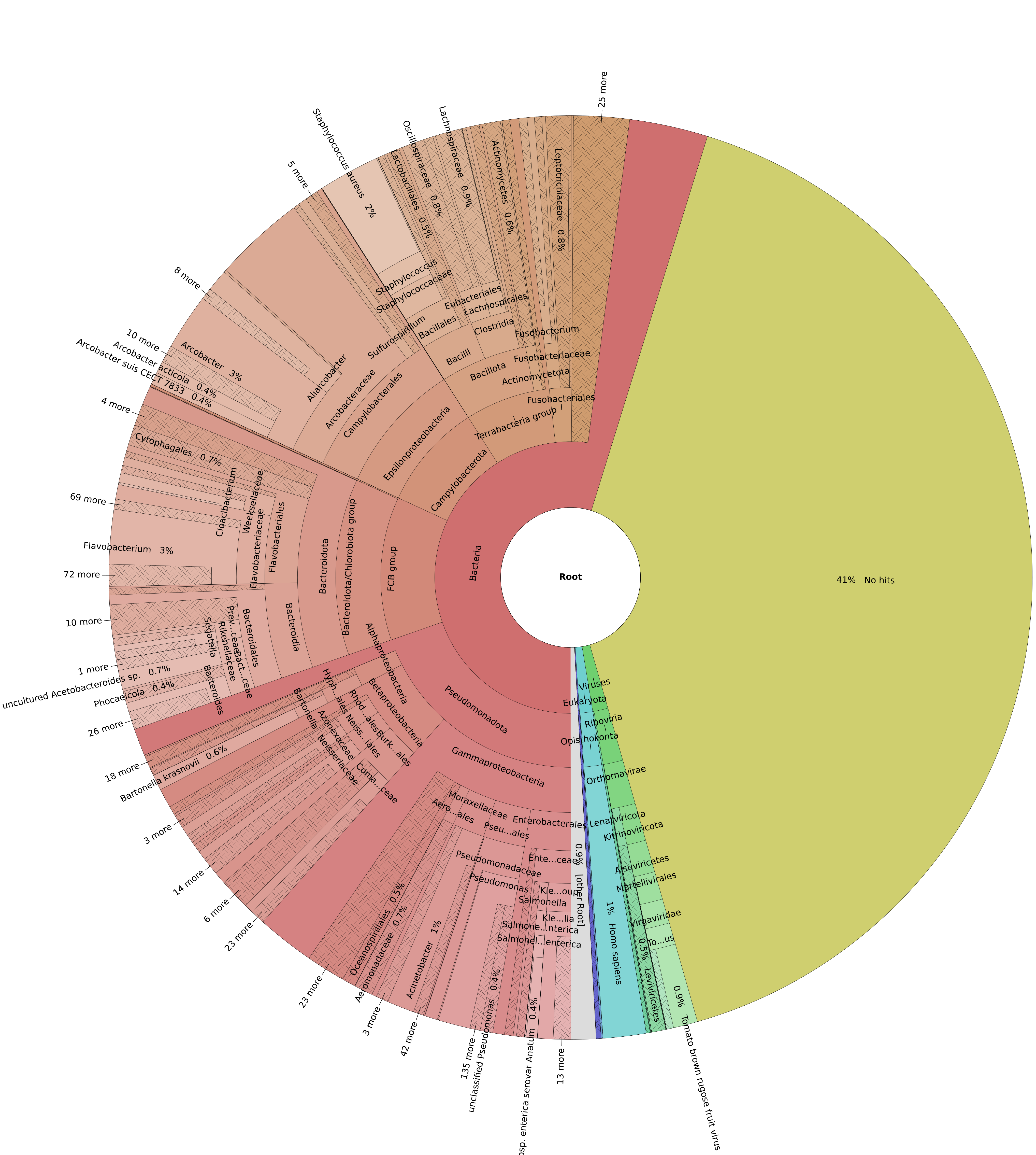}
    \vspace{-3mm}
    \caption{Metagenomic composition of the METAGENE-1 pretraining dataset, estimated via \textit{Kraken 2}~\citep{wood2019improved} sequence classification, and visualized via \textit{Krona}~\citep{ondov2011interactive}. See Figure~\ref{fig:data-snapshots-figure} for a more-detailed view.}
    \label{fig:pretraining-data-snapshot}
\end{figure}

We show an estimate of the metagenomic composition of this pretraining dataset in Figure~\ref{fig:pretraining-data-snapshot}, using the \textit{Kraken~2}~\citep{wood2019improved} sequence classification software (see Figure~\ref{fig:data-snapshots-figure} for a more-detailed view). At the highest level, this visualization shows that 55\% of reads are hits for bacteria, 2\% of reads are eukaryotes (predominantly \textit{Homo sapiens}), 2\% of reads are viruses, and 41\% of reads have \textit{no hits} and are unclassified or of unknown origin.

\subsection{Tokenization}
\label{sec:tokenization}

In developing our metagenomic foundation model, we sought a tokenization strategy that would enable high-accuracy sequence modeling, accommodate novel nucleic acid sequences, and align with best practices in modern large language models. We opted for byte-pair encoding (BPE) as our tokenization method, as it satisfies these criteria, and drawing inspiration from its successful application in recent genomic models.

BPE offers several advantages for our model. Unlike fixed-length $k$-mer tokenization, it allows for flexible token sizes, which is beneficial for capturing varying levels of genomic information, and can allow the model to adapt to different sequence patterns and structures. Moreover, BPE's ability to tokenize novel sequences is particularly valuable for modeling diverse metagenomic sequences containing unknown, varied, and possibly novel organisms. The method also has the potential to capture semantic information within a vocabulary of tokens, which can lead to more nuanced representations of genomic data.

To implement this strategy, we first trained a BPE tokenizer on a uniformly-at-random sampled subset of our pretraining dataset, comprising 2 billion base pairs. After analyzing the distribution of token sizes and considering training efficiency, we settled on a vocabulary size of 1,024 unique tokens. This vocabulary size strikes a balance between capturing sufficient genomic complexity, maintaining sufficiently long sequence lengths (based on the distribution of token sizes), and allowing for computational efficiency. Following this tokenizer training, we applied this BPE tokenizer to our entire pretraining dataset, effectively preparing it for model ingestion and training, yielding a set of $\sim$370 billion tokens ($\approx$1.69 trillion base pairs) for pretraining. We give a table showing full tokenizer details, including a list of all special tokens, in Appendix~\ref{app:tokenizer-details}.

\subsection{METAGENE-1 Architecture}
\label{sec:metagene-architecture}

For our metagenomic foundation model, we pretrain a 7-billion-parameter autoregressive language model, using a standard dense transformer architecture, similar to the architecture used in popular language models such as the GPT and Llama model families~\citep{radford2019language,llama2}.
Specifically, we implement a decoder-only style transformer with a causal language modeling objective, where the model aims to predict the next token in a sequence based on the previous tokens.

\begin{wraptable}{r}{0.38\textwidth}
\vspace{-2mm}
\footnotesize
\centering
{\begin{tabular}{l l}
\toprule
\rowcolor{Gray} \textbf{Model Details} & \textbf{METAGENE-1} \bigstrut\\
\midrule
Architecture & Llama-2-7B \bigstrut\\
Embedding Size & 4096 \bigstrut\\
Intermediate Size & 11008 \bigstrut\\
Number of Attention Heads & 32 \bigstrut\\
Number of Hidden Layers & 32 \bigstrut\\
Vocabulary Size & 1024 \bigstrut\\
Sequence Length & 512 \bigstrut\\
Normalization & RMSNorm \bigstrut\\
Regularization & $z$-loss \bigstrut\\
Position Embedding & Rotary \bigstrut\\
Bias & None \bigstrut\\
Warmup Steps & 2000 \bigstrut\\
Batch Size & 30720 \bigstrut\\
Weight Decay & 0.1 \bigstrut\\
Learning Rate Schedule & Cosine Decay \bigstrut\\
Initial Learning Rate & $6 \times 10^{-4}$ \bigstrut\\
$\beta_1$, $\beta_2$ & $0.9$, $0.95$ \bigstrut\\
\bottomrule
\end{tabular}}\\
\caption{\small METAGENE-1 architecture details.}
\label{tab:architecture-details}
\vspace{-5mm}
\end{wraptable}

This architecture choice for METAGENE-1 stands in contrast to some of the alternative approaches explored in recent genomic models, which include BERT-style bidirectional encoders~\citep{ji2021dnabert,zhou2023dnabert,zhou2024dnabert} or non-attention based architectures~\citep{nguyen2024hyenadna,nguyen2024sequence}.
Our decision to use this particular model architecture was driven by the following motivations:
\begin{enumerate}[leftmargin=7mm,itemsep=2mm,topsep=0mm] 
    \item \textit{Ecosystem}: By aligning with this widely-adopted architecture, we can take advantage of the growing ecosystem of techniques and associated implementations developed for autoregressive decoder-only transformer models. This extends to both pretraining optimizations and downstream applications in fine-tuning and inference.
    \item \textit{Infrastructure}: Given our large dataset size, this architecture allows us to leverage scalable pretraining infrastructure specifically designed for distributed training of this model type. This infrastructure has demonstrated success in recent language models, enabling efficient training on massive datasets. 
    \item \textit{Data characteristics}: The nature of our metagenomic sequence data, which primarily consists of short sequences, does not necessitate architectures designed for extremely long context lengths. This makes the transformer a suitable and efficient choice for our use case.
\end{enumerate}
We next describe some of the specific configuration details of METAGENE-1. First, the model operates with a context length of 512 tokens, which is sufficient for all of the metagenomic sequences in our pretraining dataset. For efficiency, we pack shorter sequences within this context window, a process detailed in Section~\ref{sec:context-stuffing} below. We use an attention mask which prevents attention between the distinct packed sequence reads. 
METAGENE-1 consists of 32 layers and 32 attention heads, with an embedding size of 4096 and a hidden layer size of 11008. We employ root mean square layer normalization throughout the model, with a normalization epsilon of 1e-5. These configurations result in a model with approximately 7 billion parameters in total. All architecture details are summarized in Table~\ref{tab:architecture-details}.

\section{Pretraining METAGENE-1}
\label{sec:training}

\subsection{Training Infrastructure}
\label{lref}

Our model is trained on four nodes, each equipped with 8 H100 SXM5 GPUs interconnected via Ethernet with 40 GB/s bandwidth.
This interconnect bandwidth poses a significant performance bottleneck, as it is an order of magnitude slower than NVIDIA's InfiniBand and faster Ethernet interconnects. Despite this limitation, we were able to achieve 40\% model FLOPS utilization (MFU)~\citep{palm} by employing a hybrid sharding strategy. Specifically, we use PyTorch's \texttt{HYBRID\_SHARD\_ZERO2} strategy implemented in its Fully Sharded Data Parallel (FSDP) utilities. This design choice provides the benefit of model and optimizer state sharding within each node, while practicing standard data parallelism across nodes to reduce the inter-node communication overhead. In practice, it only requires an all-reduce operation on the gradient buckets during the optimizer step.

For training, we use a global batch size of 30,720, a sequence length of 512, and a micro-batch size of 48. We observe this combination to offer the best trade-off between high MFU and reduced memory usage; it also allows us to shard the optimizer state and gradients within a single node. Further tests on fewer nodes yield MFU values of $0.51$ and $0.47$ for 1-node and 2-node setups, respectively. These results suggest that interconnect bandwidth was the main bottleneck in our training environment.

\textbf{Node failure.} During training, we experienced three node failures, one GPU failure, one network failure, and one disk failure. All failures required us to restart the training from the latest checkpoint.

\subsection{Stability}
\label{sec:training-stability}

\begin{wrapfigure}{R}{0.46\textwidth}
\centering
\vspace{-3mm}
\includegraphics[width=0.45\textwidth]{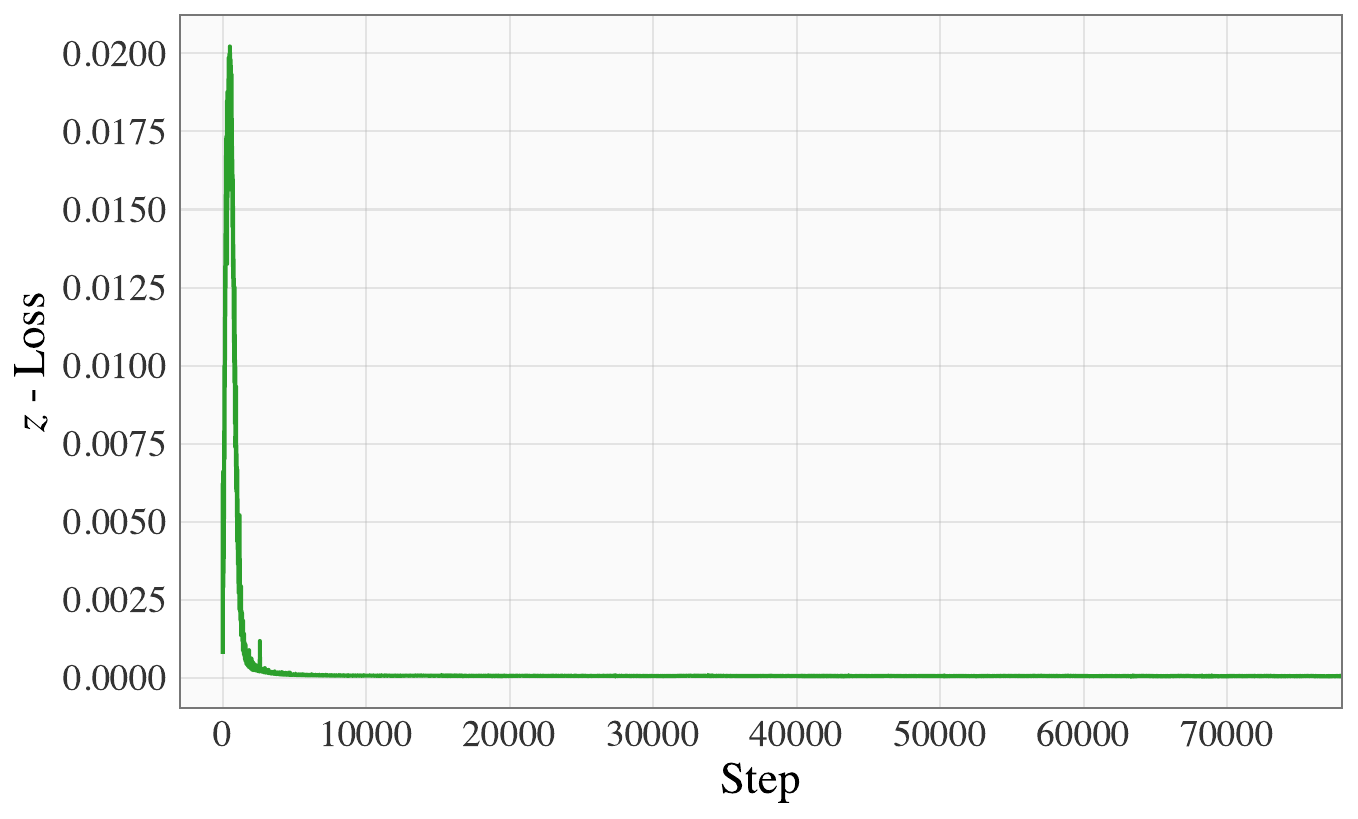}
\vspace{-1mm}
\caption{We show $z$-loss during pretraining, which aids and gives an indicator of stability.}
\label{fig:z-loss-curve}
\vspace{-2mm}
\end{wrapfigure}

Foundation model pretraining is prone to suffer from training instability, which can be more pronounced when scaling models to billions of parameters~\citep{stability}.
Such instabilities often arise during the middle or late stages of training, and are often characterized by a sudden spike in loss and/or other divergent behaviors. Failure to identify these problems can result in considerable wasted compute resources. Additionally, the characteristics of the input data have been shown to influence training stability, as highlighted by recent work in large multimodal language models~\citep{team2024chameleon}.

Given that we scaled directly from sub-billion parameters to a 7 billion parameter model, and that training on metagenomic sequences is less studied compared to natural language, we anticipated a relatively high risk of encountering stability issues. To mitigate such risks, we followed best practices from \citeauthor{stability} and implemented a variant of the \textit{z-loss}, referred to as \textit{max-z-loss}, introduced by \citeauthor{baichuan2} with a coefficient of 2e-4.
We opted against the recommendation of QK-layer normalization~\citep{team2024chameleon} to preserve the Llama architecture and leverage optimized inference pipelines.

During training, we monitored the norms of the language model head, the query, key, and value outputs, as well as the gradient norms. \citeauthor{stability} empirically shows that a significant increase in any of these metrics may signify potential instability, allowing us to intervene early by restarting the training. Fortunately, no stability issues were observed, and these metrics remained consistent throughout the training process.

\vspace{-1mm}
\subsection{Context Stuffing}
\label{sec:context-stuffing}

A significant portion of our dataset contains sequences with fewer tokens than our model's context length. To optimize compute efficiency and avoid wasting resources on padding tokens, we pack the sequence dimension with multiple samples, where applicable. We modify the attention mask to ensure that tokens from different samples cannot attend to one another.
This is implemented using the variable length function in \textit{FlashAttention-2}\footnote{Named function \texttt{flash\_attn\_varlen\_func} in the \textit{FlashAttention-2} Python package.}~\citep{fa2} which avoids materializing the full mask, which would have been inefficient.

\begin{figure}[t!]
    \centering
    \hspace{-2mm}
    \includegraphics[width=0.47\textwidth]{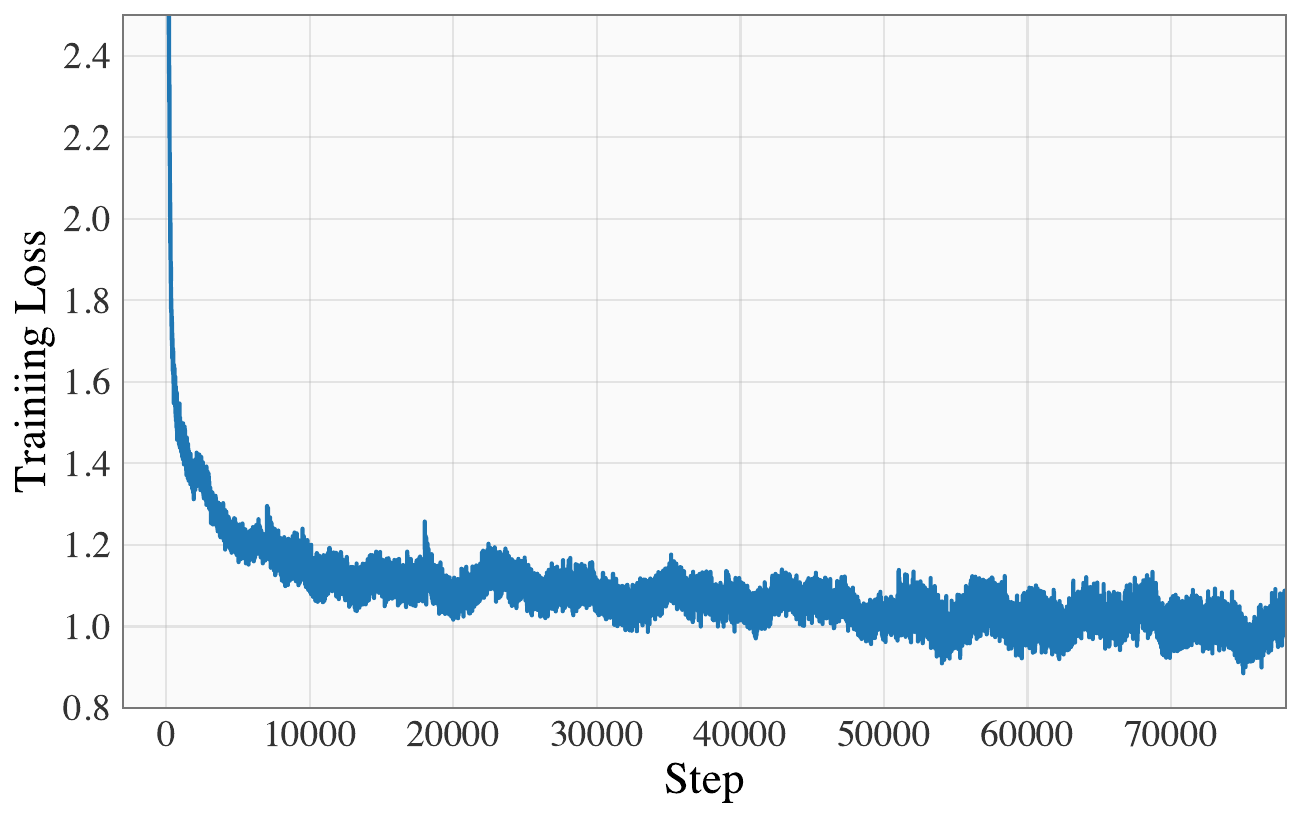}
    \hspace{6mm}
    \includegraphics[width=0.47\textwidth]{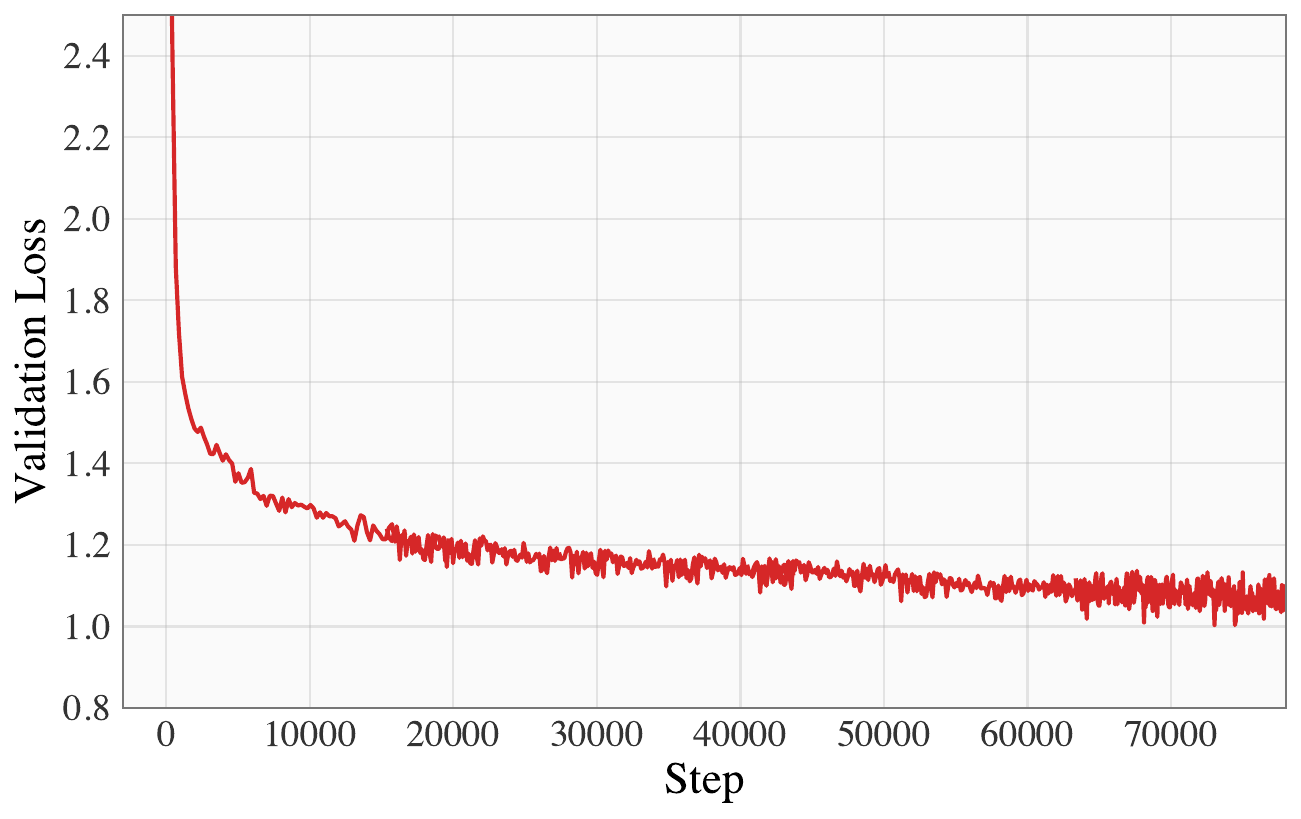}
    \caption{METAGENE-1 loss curves during pretraining. We show training loss (left), and validation loss on a held out metagenomic sample (right).}
    \label{fig:loss-curves}
\end{figure}

\vspace{-1mm}
\subsection{Continual Pretraining}
\label{sec:second-stage}

After the initial stage of pretraining is complete, we carry out a second stage of pretraining which constitutes about 9\% of our total number of pretraining tokens. In this second stage of training, we extend our dataset to a broader distribution of genomic sequences relative to our original metagenomic distribution, and we follow practices for continual learning, such as annealing the learning rate both to enact a \textit{warmup} period (\textit{i.e.}, a linear ramp up to account for the shifted data distribution), and a \textit{cooldown} period (\textit{i.e.}, a ramp down of the learning rate at the end of training for improved performance \citep{hagele2024scaling}).

The modified training distribution aims to allow for us to maintain performance on metagenomic tasks, such as metagenomic embedding and classification, while also achieving improved performance on a broader set of genomic tasks (\textit{i.e.}, tasks involving non-metagenomic data). For this, we sample sequences from the dataset provided by~\citet{zhou2023dnabert}, which includes genomic sequences from known organisms---both from human genomes and a curated selection of genomes from multiple species (\textit{e.g.}, fungi, mammalian, invertebrate, bacteria)---and shuffle it into our metagenomic reads at a 1:8 ratio.

\section{Empirical Results}
\label{sec:exp}

\subsection{Pretraining Performance}
\label{sec:pretraining-performance}

As an initial analysis of METAGENE-1, in Figure~\ref{fig:loss-curves}, we show two loss curves generated over the course of pretraining.
On the left, we show the training loss over one epoch of our 1.5-trillion-base-pair pretraining dataset. On the right, we show the validation loss, computed on a held-out portion of our metagenomic dataset.
In the training curve we note that there are slight systematic oscillations over the course of training, which occur due to pseudo-random data shuffling (implemented for efficiency reasons); however, these do not appear in our validation loss curve.

\vspace{-2mm}
\subsection{Pathogen Detection Benchmark}
\label{sec:pathogen-detection}

Our initial experiments evaluate METAGENE-1's reliability in detecting human pathogens. To this end, we construct four datasets with binary labels, aiming to classify human pathogens versus non-pathogens. These datasets are constructed from four distinct sequencing deliveries, which are excluded from our pretraining data. For each delivery, we extract two sets of sequencing reads: pathogen and non-pathogen. Pathogen reads are defined as a subset of sequencing reads meeting two criteria: (1) Kraken 2 \citep{wood2019improved}\footnote{We use the 2024-06 Standard Database for identification. $^6$We use the 2024-06 GenBank release available at \url{https://www.ncbi.nlm.nih.gov/genbank/}.} identifies at least one hit on a $k$-mer associated with a human-infecting virus, and (2) the read aligns with a human-infecting virus genome in GenBank\footnotemark{}. The sub-tasks in this pathogen detection benchmark represent different deliveries, which vary by collection location, sequencing pipeline, date, or a combination of these factors. Each dataset contains 1,600 training samples and 2,000 test samples. We intentionally use a small training set to mimic real-world scenarios where rare human pathogens are expensive to identify.

\begin{table}[h!]
\centering
\resizebox{0.96\textwidth}{!}{
\begin{tabular}{lcccc|c}
\toprule
\rowcolor{Cornsilk}
& \textbf{DNABERT-2} & \textbf{DNABERT-S} & \textbf{NT-2.5b-Multi} & \textbf{NT-2.5b-1000g} & \textbf{METAGENE-1} \\
\midrule
\rowcolor{Gray} \textbf{\textsc{Pathogen-Detect (avg.)}}        & 87.92    & 87.02 &  82.43 & 79.02                & \textbf{92.96}                \\
\midrule
{\small \textbf{\textsc{Pathogen-Detect-1}}}                    & {\small 86.73} & {\small 85.43} & {\small 83.80} & {\small 77.52}                & {\small \textbf{92.14}}            \\
{\small \textbf{\textsc{Pathogen-Detect-2}}}                    & {\small 86.90} & {\small 85.23} & {\small 83.53} & {\small 80.38}                & {\small \textbf{90.91}}            \\
{\small \textbf{\textsc{Pathogen-Detect-3}}}                    & {\small 88.30} & {\small 89.01} & {\small 82.48} & {\small 79.83}                & {\small \textbf{93.70}}            \\
{\small \textbf{\textsc{Pathogen-Detect-4}}}                    & {\small 89.77} & {\small 88.41} & {\small 79.91} & {\small 78.37}                & {\small \textbf{95.10}}            \\
\bottomrule
\end{tabular}
}
\caption{Results on the Pathogen Detection benchmark. The metric used for all evaluations is MCC. The header row reports macro-averaged performance metrics. See Section~\ref{sec:pathogen-detection} for details.}
\label{tab:pathogen-detection}
\end{table}

We evaluate the performance of METAGENE-1 and other genomic foundation models on the pathogen detection datasets, measured using the Matthews correlation coefficient (MCC). All models were trained with a consistent set of hyperparameters: DNABERT~\citep{zhou2024dnabert} variants undergo full-model fine-tuning, while Nucleotide Transformer (NT)~\citep{dalla2023nucleotide} variants and METAGENE-1 are fine-tuned using low-rank adapters (LoRA)~\citep{hu2021lora}. For sequence-level classification, we use the built-in pooler for DNABERT and NT models provided in HuggingFace Transformers~\citep{wolf2019huggingface}, and use mean-pooled representations for METAGENE-1. Additional experimental details can be found in~\Cref{app:pathogen-details}.

As shown in \Cref{tab:pathogen-detection}, METAGENE-1 consistently outperforms all other models across the Pathogen Detection benchmark, with gains ranging from approximately 3 to 17 MCC points over the strongest competing models. These results highlight METAGENE-1's strong performance in pathogen detection tasks, particularly in scenarios with diverse sequencing conditions or delivery pipelines.

\subsection{Genomic Embedding Benchmark}
\label{sec:gene-mteb}

\begin{table}[t!]
\centering
\resizebox{0.9\textwidth}{!}{
\begin{tabular}{lcccc|c}
\toprule
\rowcolor{Cornsilk}
& \textbf{DNABERT-2} & \textbf{DNABERT-S} & \textbf{NT-2.5b-Multi} & \textbf{NT-2.5b-1000g} & \textbf{METAGENE-1} \\
\midrule
\rowcolor{Gray} \textbf{\textsc{Human-Virus (avg.)}}            & 0.564            & 0.570            & 0.675                       & 0.710                & \textbf{0.775}            \\
\midrule
{\small \textbf{\textsc{Human-Virus-1}}}                    & {\small 0.594}            & {\small 0.605}            & {\small 0.671}                       & {\small 0.721}                & {\small \textbf{0.828}}            \\
{\small \textbf{\textsc{Human-Virus-2}}}                    & {\small 0.507}            & {\small 0.510}            & {\small 0.652}                       & {\small 0.624}                & {\small \textbf{0.742}}            \\
{\small \textbf{\textsc{Human-Virus-3}}}                    & {\small 0.606}            & {\small 0.612}            & {\small 0.758}                       & {\small 0.740}                & {\small \textbf{0.835}}            \\
{\small \textbf{\textsc{Human-Virus-4}}}                    & {\small 0.550}            & {\small 0.551}            & {\small 0.620}                       & {\small \textbf{0.755}}                & {\small 0.697}            \\
\midrule
\rowcolor{Gray} \textbf{\textsc{HMPD (avg.)}}          & 0.397            & 0.403            & 0.449                       & 0.451                & \textbf{0.465}            \\
\midrule
{\small \textbf{\textsc{HMPD-single}}}                    & {\small 0.292}            & {\small 0.293}            & {\small 0.285}                       & {\small 0.292}                & {\small \textbf{0.297}}            \\
{\small \textbf{\textsc{HMPD-disease}}}            & {\small 0.480}            & {\small 0.486}            & {\small 0.498}                       & {\small 0.489}                & {\small \textbf{0.542}}            \\
{\small \textbf{\textsc{HMPD-sex}}}                & {\small 0.366}            & {\small 0.367}            & {\small 0.487}                       & {\small 0.476}                & {\small \textbf{0.495}}            \\
{\small \textbf{\textsc{HMPD-source}}}             & {\small 0.451}            & {\small 0.465}            & {\small 0.523}                       & {\small \textbf{0.545}}                & {\small 0.526}            \\
\midrule
\rowcolor{Gray} \textbf{\textsc{HVR (avg.)}}           & 0.479            & 0.479            & 0.546                       & 0.524                & \textbf{0.550}            \\
\midrule
{\small \textbf{\textsc{HVR-p2p}}}                 & {\small 0.548}            & {\small 0.550}            & {\small 0.559}                       & {\small \textbf{0.650}}                & {\small 0.466}            \\
{\small \textbf{\textsc{HVR-s2s-align}}}           & {\small 0.243}            & {\small 0.241}            & {\small 0.266}                       & {\small \textbf{0.293}}                & {\small 0.267}            \\
{\small \textbf{\textsc{HVR-s2s-small}}}           & {\small 0.373}            & {\small 0.372}            & {\small 0.357}                       & {\small 0.371}                & {\small \textbf{0.467}}            \\
{\small \textbf{\textsc{HVR-s2s-tiny}}}            & {\small 0.753}            & {\small 0.753}            & {\small 1.000}                       & {\small 0.782}                & {\small \textbf{1.000}}            \\
\midrule
\rowcolor{Gray} \textbf{\textsc{HMPR (avg.)}}          & 0.347            & 0.351            & 0.348                       & 0.403                & \textbf{0.476}            \\
\midrule
{\small \textbf{\textsc{HMPR-p2p}}}                & {\small 0.566}            & {\small \textbf{0.580}}            & {\small 0.471}                       & {\small 0.543}                & {\small 0.479}            \\
{\small \textbf{\textsc{HMPR-s2s-align}}}          & {\small 0.127}            & {\small 0.129}            & {\small 0.144}                       & {\small \textbf{0.219}}                & {\small 0.140}            \\
{\small \textbf{\textsc{HMPR-s2s-small}}}          & {\small 0.419}            & {\small 0.421}            & {\small 0.443}                       & {\small \textbf{0.459}}                & {\small 0.432}            \\
{\small \textbf{\textsc{HMPR-s2s-tiny}}}           & {\small 0.274}            & {\small 0.274}            & {\small 0.332}                       & {\small 0.391}                & {\small \textbf{0.855}}            \\
\midrule
\midrule
\rowcolor{Gray} \textbf{\textsc{Global Average}}                 & 0.475            & 0.479            & 0.525                       & 0.545                & \textbf{0.590}            \\
\bottomrule
\end{tabular}
}
\caption{Results on the Genomic Embedding (Gene-MTEB) benchmark. See Section~\ref{sec:gene-mteb} for details.}
\vspace{-3mm}
\label{tab:gene-mteb-results}
\end{table}

Next, we assess METAGENE-1's ability to generate high-quality representations in a zero-shot manner. These representations are crucial for lightweight development of predictive models using a frozen foundation model~\citep[\textit{inter alia}]{devlin2018bert,karpukhin2020dense}.
They enhance interpretability by enabling sparse autoencoders to produce semantically meaningful encodings~\citep{bricken2023monosemanticity,gao2024scaling}.
Additionally, they are vital for anomaly detection methods that rely on them for effective modeling~\citep{yang2024ad}. Drawing inspiration from MTEB~\citep{muennighoff2022mteb}, we introduce a large-scale genomics embedding benchmark, termed Gene-MTEB, to advance the development of robust genomics representations.

For this benchmark, we curate eight classification tasks (Human-Virus-1-4, MHPD-single, HMPD-disease, HMPD-source, HMPD-sex),
and eight clustering tasks (HVR-p2p, HVR-s2s-align, HVR-s2s-small, HVR-s2s-tiny, HMPR-p2p,HMPR-s2s-align, HMPR-s2s-small, HMPR-s2s-tiny). Datasets for these tasks are sourced from the Human Microbiome Project~\citep{peterson2009nih}, and held-out portions of our metagenomic dataset. Details and access to all benchmark datasets are provided on the project HuggingFace page. All classification tasks carry out logistic regression on top of embeddings and all clustering tasks carry out mini-batch $k$-means. Embeddings for all models are accessed via mean pooling on the last hidden state. 

Results on Gene-MTEB are shown in Table~\ref{tab:gene-mteb-results}. Here, \textit{accuracy} is shown for classification and \textit{V-measure} for clustering tasks. We find that METAGENE-1 shows strong embedding performance across the board, and in particular for Human-Virus datasets, scoring over 6 points above all other models. Continual training with representation learning objectives, such as contrastive losses, could further enhance its embedding quality beyond its current LM-based pretraining.

\subsection{Genome Understanding Evaluation Benchmark}
\label{sec:fine-tuning-performance}

\begin{table}[t!]
\centering
\resizebox{0.86\textwidth}{!}{
\hspace{-2mm}
\begin{tabular}{cccccc|c}
    \toprule
    \rowcolor{Cornsilk}
   & \textbf{CNN} & \textbf{HyenaDNA} & \textbf{DNABERT} & \textbf{NT-2.5B-Multi} & \textbf{DNABERT-2} & \textbf{METAGENE-1} \\
   \midrule
\rowcolor{Gray} \textbf{\textsc{TF-Mouse (avg.)}} & 45.3 & 51.0 & 57.7 & 67.0 & 68.0 & \textbf{71.4} \\
\midrule
{\small \textbf{\textsc{0}}} & {\small 31.1} & {\small 35.6} & {\small 42.3} & {\small \textbf{63.3}} & {\small 56.8} & {\small 61.5} \\
{\small \textbf{\textsc{1}}} & {\small 59.7} & {\small 80.5} & {\small 79.1} & {\small 83.8} & {\small \textbf{84.8}} & {\small 83.7} \\
{\small \textbf{\textsc{2}}} & {\small 63.2} & {\small 65.3} & {\small 69.9} & {\small 71.5} & {\small 79.3} & {\small \textbf{83.0}} \\
{\small \textbf{\textsc{3}}} & {\small 45.5} & {\small 54.2} & {\small 55.4} & {\small 69.4} & {\small 66.5} & {\small \textbf{82.2}} \\
{\small \textbf{\textsc{4}}} & {\small 27.2} & {\small 19.2} & {\small 42.0} & {\small 47.1} & {\small \textbf{52.7}} & {\small 46.6} \\
\midrule
\rowcolor{Gray} \textbf{\textsc{TF-Human (avg.)}} & 50.7 & 56.0 & 64.4 & 62.6 & \textbf{70.1} & 68.3 \\
\midrule
{\small \textbf{\textsc{0}}} & {\small 54.0} & {\small 62.3} & {\small 68.0} & {\small 66.6} & {\small \textbf{72.0}} & {\small 68.9} \\
{\small \textbf{\textsc{1}}} & {\small 63.2} & {\small 67.9} & {\small 70.9} & {\small 66.6} & {\small \textbf{76.1}} & {\small 70.8} \\
{\small \textbf{\textsc{2}}} & {\small 45.2} & {\small 46.9} & {\small 60.5} & {\small 58.7} & {\small \textbf{66.5}} & {\small 65.9} \\
{\small \textbf{\textsc{3}}} & {\small 29.8} & {\small 41.8} & {\small 53.0} & {\small 51.7} & {\small \textbf{58.5}} & {\small 58.1} \\
{\small \textbf{\textsc{4}}} & {\small 61.5} & {\small 61.2} & {\small 69.8} & {\small 69.3} & {\small 77.4} & {\small \textbf{77.9}} \\
\midrule
\rowcolor{Gray} \textbf{\textsc{EMP (avg.)}} & 37.6 & 44.9 & 49.5 & 58.1 & 56.0 & \textbf{66.0} \\
\midrule
{\small \textbf{\textsc{H3}}} & {\small 61.5} & {\small 67.2} & {\small 74.2} & {\small 78.8} & {\small 78.3} & {\small \textbf{80.2}} \\
{\small \textbf{\textsc{H3K14ac}}} & {\small 29.7} & {\small 32.0} & {\small 42.1} & {\small 56.2} & {\small 52.6} & {\small \textbf{64.9}} \\
{\small \textbf{\textsc{H3K36me3}}} & {\small 38.6} & {\small 48.3} & {\small 48.5} & {\small 62.0} & {\small 56.9} & {\small \textbf{66.7}} \\
{\small \textbf{\textsc{H3K4me1}}} & {\small 26.1} & {\small 35.8} & {\small 43.0} & {\small 55.3} & {\small 50.5} & {\small \textbf{55.3}} \\
{\small \textbf{\textsc{H3K4me2}}} & {\small 25.8} & {\small 25.8} & {\small 31.3} & {\small 36.5} & {\small 31.1} & {\small \textbf{51.2}} \\
{\small \textbf{\textsc{H3K4me3}}} & {\small 20.5} & {\small 23.1} & {\small 28.9} & {\small 40.3} & {\small 36.3} & {\small \textbf{58.5}} \\
{\small \textbf{\textsc{H3K79me3}}} & {\small 46.3} & {\small 54.1} & {\small 60.1} & {\small 64.7} & {\small 67.4} & {\small \textbf{73.0}} \\
{\small \textbf{\textsc{H3K9ac}}} & {\small 40.0} & {\small 50.8} & {\small 50.5} & {\small 56.0} & {\small 55.6} & {\small \textbf{65.5}} \\
{\small \textbf{\textsc{H4}}} & {\small 62.3} & {\small 73.7} & {\small 78.3} & {\small 81.7} & {\small 80.7} & {\small \textbf{82.7}} \\
{\small \textbf{\textsc{H4ac}}} & {\small 25.5} & {\small 38.4} & {\small 38.6} & {\small 49.1} & {\small 50.4} & {\small \textbf{61.7}} \\
\midrule
\rowcolor{Gray} \textbf{\textsc{PD (avg.)}} & {77.1} & {35.0} & {84.6} & \textbf{88.1} & {84.2} & {82.3} \\
\midrule
{\small \textbf{\textsc{All}}} & {\small 75.8} & {\small 47.4} & {\small 90.4} & {\small \textbf{91.0}} & {\small 86.8} & {\small 86.0} \\
{\small \textbf{\textsc{No-TATA}}} & {\small 85.1} & {\small 52.2} & {\small 93.6} & {\small 94.0} & {\small \textbf{94.3}} & {\small 93.7} \\
{\small \textbf{\textsc{TATA}}} & {\small 70.3} & {\small 5.3} & {\small 69.8} & {\small \textbf{79.4}} & {\small 71.6} & {\small 67.4} \\
\midrule
\rowcolor{Gray} \textbf{\textsc{CPD (avg.)}} & {62.5} & {48.4} & \textbf{73.0} & {71.6} & {70.5} & {69.9} \\
\midrule
{\small \textbf{\textsc{All}}} & {\small 58.1} & {\small 37.0} & {\small \textbf{70.9}} & {\small 70.3} & {\small 69.4} & {\small 66.4} \\
{\small \textbf{\textsc{No-TATA}}} & {\small 60.1} & {\small 35.4} & {\small 69.8} & {\small \textbf{71.6}} & {\small 68.0} & {\small 68.3} \\
{\small \textbf{\textsc{TATA}}} & {\small 69.3} & {\small 72.9} & {\small \textbf{78.2}} & {\small 73.0} & {\small 74.2} & {\small 75.1} \\
\midrule
\textbf{\textsc{SSD}} & {76.8} & {72.7} & {84.1} & \textbf{89.3} & {85.0} & {87.8} \\
\midrule
\textbf{\textsc{COVID}} & {22.2} & {23.3} & {62.2} & \textbf{73.0} & {71.9} & {72.5} \\
\midrule
\midrule
\rowcolor{Gray} \textbf{\textsc{Global Win \%}} & {0.0} & {0.0} & {7.1} & {21.4} & {25.0} & \textbf{46.4} \\
\bottomrule
\end{tabular}
}
    \vspace{2mm}
    \caption{Results on the Genome Understanding Evaluation (GUE) benchmark. Non-METAGENE-1 results are adapted from \citet{zhou2023dnabert}. The metric used for all evaluations is MCC, except for the COVID task, which uses F1 score. The header rows report macro-averaged performance metrics. The final row shows \textit{Global Win \%}, \textit{i.e.}, the percentage of tasks in which a given method achieves top score under the associated metric.}
    \label{tab:gue-results}
\end{table}

We now investigate the viability of METAGENE-1 as a general-purpose foundation model. Importantly, we aim to assess its performance on nucleotide sequences sampled from a diverse array of species. One such example is long-sequence full-animal-genome datasets. In many prior genomic sequence models' pretraining datasets, this type of genomic data is found in abundance~\citep{dalla2023nucleotide, ji2021dnabert, nguyen2024hyenadna, zhou2023dnabert}.
As a pilot study, we perform fine-tuning experiments on the Genome Understanding Evaluation (GUE) benchmark~\citep{zhou2023dnabert}, which comprises 28 sequence-level classification tasks curated from this type of genomics data.

Following~\Cref{sec:pathogen-detection}, we fine-tune low-rank adapters (LoRA)~\citep{hu2021lora} and a linear classification head that projects average-pooled representations from the last hidden layer to the class logits. This setup is aimed to emulate downstream users with a limited compute budget. For each experiment, we perform a grid search over linearly spaced learning rates from 1e-4 to 1e-3 and select LoRA modules from query-value and query-key-value-dense combinations. We fix all other hyperparameters and select the best configuration based on validation performances. Additional details on training hyperparameters can be found in \Cref{app:gue-details}. Following the metrics selected in~\citeauthor{zhou2023dnabert}, we report Matthews correlation coefficient (MCC) on all but the COVID task, which instead uses the F1 score. 

In \Cref{tab:gue-results}, we present METAGENE-1's performance on the GUE benchmark. Our findings show that METAGENE-1 outperforms or remains competitive with state-of-the-art foundation models specializing in multi-species genomics prediction, achieving a top score on 13 out of 28 GUE subtasks (compared with DNABERT-2, the second highest scoring model, that achieves a top score on 7 out of 28 subtasks). Notably, METAGENE-1 excels in Epigenetic Marks Prediction (EMP) tasks but shows room for improvement in (Core) Promoter Detection (PD/CPD). We attribute this to limitations in the pre-training data mixture, and believe that a more tailored pre-training dataset could potentially enhance METAGENE-1's performance in this area.

\begin{figure}[h]
    \centering
    \includegraphics[width=1\linewidth]{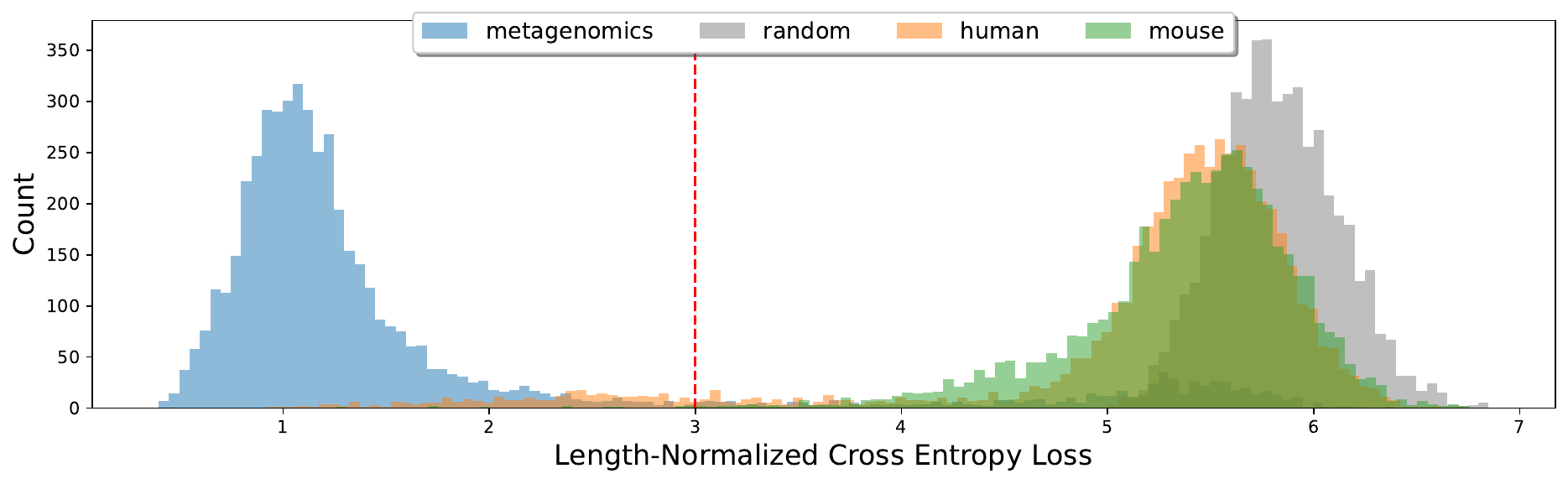}
    \caption{Distribution of the length-normalized cross entropy loss across all datasets, given by METAGENE-1.}
    \label{fig:loss-histogram}
\end{figure}

\begin{table}[h]
    \centering
    \begin{tabular}{|l|c|c|c|}
      \hline
      \textbf{Group} & \textbf{F1} & \textbf{Loss (Std. Err)} & \textbf{Tokenized Seq Len (Std. Dev)} \\
      \hline
      \textbf{Metagenomics} & - & 1.24 (1.31) & 24.91 (3.35) \\
      \textbf{Random} & 0.91 & 5.83 (0.29) & 27.16 (1.32)   \\
      \textbf{Human} & 0.94 & 5.22 (0.22) &  27.29 (1.33) \\
      \textbf{Mouse} & 0.91 & 5.38 (0.54) &  27.2 (1.34) \\
      \hline
    \end{tabular}
    \vspace{2mm}
    \caption{OOD detection performance between metagenomics sequences and other data sources. }
    \label{tab:ood-results}
\end{table}

\subsection{Anomaly Detection from Wastewater}
\label{sec:anomaly-detection}

Our final experiment aims to show the feasibility of METAGENE-1 to detect out-of-distribution (OOD) data at scale, as it serves as a primer for reliable anomaly detection from wastewater samples. In this early study, we sample 5000 sequences from, respectively, our metagenomics pretraining data, the mouse and human genomes from the GUE dataset, as well as  \textit{uniform random} sequences as a control group. All sequences are truncated to 100 base pairs in accordance with the sequence lengths from the GUE dataset. As a baseline, we implement a threshold-based anomaly detector, which classifies samples with length-normalized cross entropy losses below a certain threshold as non-anomalies, and \textit{vice versa}. We select a threshold of 3 based on our observations from the validation curve in \Cref{fig:loss-curves}. Note that this anomaly detection study is performed using a checkpoint of METAGENE-1 that has only been pretrained on metagenomic data (\textit{i.e.}, without second-stage training).

\Cref{fig:loss-histogram} indicates a clear separation between metagenomics sequences and other data sources. The in-distribution data behaves within our expectation; the human and mouse genomic data both attain a similar mode and spread, and their loss distributions are more similar to that of random sequences, compared to our in-distribution data. \Cref{tab:ood-results} reports numerical results of our OOD detection tests. METAGENE-1 achieves strong performance for separating metagenomics sequences from other data sources.

\section{Safety Considerations}
\label{sec:safety}

Metagenomic foundation models like METAGENE-1 demonstrate improved capabilities on tasks that can aid in biosurveillance, genomic anomaly detection, and pandemic monitoring. While still relatively small in scale compared with many modern language models, METAGENE-1 shows state-of-the-art results on benchmarks and enables potential downstream uses. However, these capabilities merit careful consideration of safety and must be balanced against potential risks. This category of genomic model---and especially, future larger variants of it---could pose risks to human health and safety by enabling harmful applications, such as the design of novel pathogenic DNA sequences or synthetic genetic materials. These potential abuses were considered when deciding to open source METAGENE-1. The final decision was based on weighing the beneficial applications, such as pandemic preparedness, against the potential for misuse. Based on our safety considerations, which we outline below, we believe that the current iteration of METAGENE-1 poses minimal risk, and its release is justified by its significant positive potential. However, we also recognize and discuss the need for careful safety considerations before open sourcing increasingly capable models of this type.

\textbf{Relation to other open source genomic models.}
METAGENE-1 is a genomic foundation model that builds upon a lineage of similar open-source efforts, such as NucleotideTransformer~\citep{dalla2023nucleotide}, DNABERT~\citep{ji2021dnabert}, HyenaDNA~\citep{nguyen2024hyenadna}, Evo~\citep{nguyen2024sequence}, and more. At 7 billion parameters, METAGENE-1 matches the largest of these existing models. The key distinction of METAGENE-1 lies in the model’s training data: a highly diverse set of metagenomic sequences derived from wastewater, with a focus on the human microbiome. This dataset, comprising short uncurated sequences from tens of thousands of species, allows METAGENE-1 to excel at representing the complexities of microbial and viral diversity in metagenomic samples, providing unique advantages in biosurveillance applications. Similar to other genomic foundation models, and unlike large language models, these models alone do not possess significant reasoning or control capabilities (given that complex control instructions cannot easily be provided via input context, which is restricted to genomic sequences).

\textbf{Tailored for detection, not design.} METAGENE-1 was specifically designed for anomaly detection in metagenomic data, not for complex genomic design tasks. The training data, model architecture, and task design are geared toward detecting and classifying anomalies in short sequences of a few hundred base pairs. Notably, all metagenomic data used in pretraining METAGENE-1 consist exclusively of sequences ranging from 100 to 300 base pairs. Unlike large genomic models focused on longer sequence generation, METAGENE-1’s capabilities are tailored to analyzing these short metagenomic reads. Its architectural constraints, including a maximum context length of 512 tokens, further limit its applicability to sequence design tasks. These design decisions ensure that the model’s primary utility lies in detecting pathogens and monitoring biosurveillance trends, rather than enabling misuse in synthetic biology.

\textbf{Pros and cons of open source.} Open sourcing a model of this type is a balance between the potential for help and harm. In the case of METAGENE-1, we believe that open source is net positive for research in the area of anomaly detection for pathogen monitoring. We hope that the availability of this model can have a positive impact on facilitating safety research, a prospect that we discuss in \Cref{sec:discussion}. Nonetheless, we recognize the importance of caution when releasing models in this domain. For future iterations of pathogen-detection models with improved capabilities, we believe strongly in (and we ourselves are are committed to) thoroughly evaluating the safety and potential for misuse before an open source release. Larger-scale models, in particular, present additional risks, and we advocate for rigorous safety assessments in determining whether such models should be released publicly. By prioritizing careful oversight and responsible scaling, we aim to mitigate risks while maximizing the benefits of this technology for public health and biosurveillance.

\section{Discussion, Limitations, Conclusion}
\label{sec:discussion}

We have reported our current progress on pretraining and evaluating METAGENE-1, the first large-scale foundation model pretrained on metagenomic sequences. We detail our dataset construction, model training, and fine-tuning procedure to facilitate open-science research. Additionally, we open-source our training code and model checkpoints. 

Our downstream performance on genomic benchmarks indicates the potential of METAGENE-1 as a general-purpose foundation model. Our results also indicate that METAGENE-1 benefits from continual pretraining on a diverse mixture of data sources in addition to metagenomic data (at least for tasks similar to these genomic benchmarks). We are continuing to actively explore this direction, through incorporating additional human reference genomes and multi-species genomic datasets in our metagenomic pretraining data.

\textbf{Limitations.} METAGENE-1 is pretrained on a dataset consisting primarily of wastewater metagenomics and multi-species genomic sequences, making it well-suited for downstream tasks within this distribution. However, like many foundation models, it requires additional fine-tuning to achieve optimal performance for specific applications. Additionally, the pretraining data predominantly consist of short metagenomic sequencing reads, limiting the model's performance to contexts involving shorter metagenomics inputs. This may restrict its effectiveness for tasks involving long-read or full-genome data, where long-sequence models may be necessary~\citep{nguyen2024sequence,nguyen2024hyenadna}.

\textbf{Future directions.} There are many potential avenues for future research. An area that we are particularly excited about concerns the \textit{understanding} of genomic foundation models. While a great deal of prior work has studied the mechanistic interpretability of language models~\citep{wang2022interpretability,hanna2024does,conmy2023towards,syed2023attribution}, their extensions beyond language and vision have been limited. Future work could systematize approaches to mechanistic interpretability in genomics by leveraging sparse autoencoders (SAEs)~\citep{bricken2023monosemanticity,gao2024scaling,lieberum2024gemma} to identify biologically meaningful features, employing attribution methods to trace model predictions to genomic regions~\citep{koo2020interpreting,tseng2020fourier,majdandzic2022selecting}, and developing new tools for probing model representations using task-specific datasets~\citep{conneau2018you,hewitt2019designing}. A better understanding of these models would not only advance their reliability but also help mitigate risks, such as inadvertently generating or propagating harmful genomic sequences. 

Finally, we are actively developing a standardized evaluation suite consisting of classification, embedding, out-of-distribution detection, and pandemic monitoring tasks for metagenomics sequences. We hope our effort can facilitate objective evaluation of METAGENE-1 and future metagenomic models, and we invite both domain experts and the machine learning community to contribute to this research.

\paragraph{Acknowledgements}
We thank Prime Intellect for computing support, and the Nucleic Acid Observatory for metagenomic data resources.
In particular, we thank Marc Johnson, Clayton Rushford, and Jason Rothman for metagenomic sequencing support that produced the data for this project.
W.N. would like to thank Victor Miller and Mark Schulze for helpful discussions and feedback on this project.
O.L. would like to thank Zhihan Zhou for helpful discussions and insights on the GUE benchmark.

\bibliography{main}
\newpage
\appendix
\newpage

\appendix

\section*{\hspace{-4mm} \centering Appendix}
\vspace{3mm}

\section{Additional Details on the Metagenomic Pretraining Dataset}
\label{app:additional-data-details}

\begin{figure}[t]
    \centering
    \vspace{-4mm}
    \includegraphics[width=0.88\textwidth]{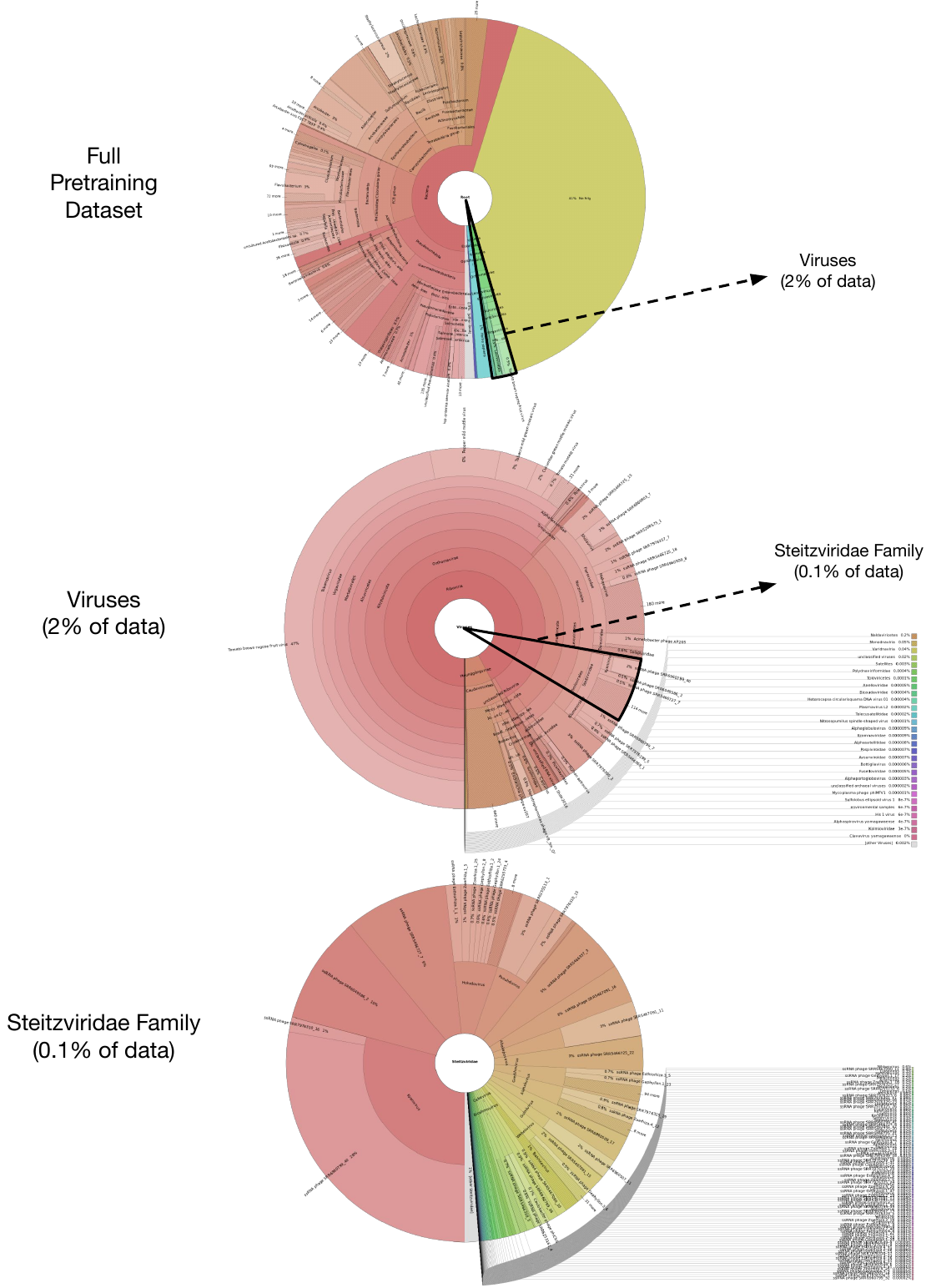}
    \caption{A visualization of the composition of metagenomic information contained in our pretraining dataset, based on \textit{Kraken 2} metagenomic sequence classification hits \citep{wood2019improved}. We first show the full pretraining dataset distribution (top), and then as an example show the distribution of viruses (middle), and finally the distribution of the Steitzviridae family of viruses (bottom).}
    \label{fig:data-snapshots-figure}
\end{figure}

In Figure~\ref{fig:data-snapshots-figure}, we show a visualization of (a relatively small subset of) the composition of metagenomic information contained in our pretraining dataset. This composition is estimated through the \textit{Kraken~2} metagenomic sequence classification software~\citep{wood2019improved}, which gives taxonomic hits for reads in our pretraining set (where taxonomic classification is performed using exact $k$-mer matches). We show three plots in Figure~\ref{fig:data-snapshots-figure}: first, the full pretraining dataset distribution (top); then, an example subset of this showing the distribution of viruses (middle); and finally, an example subset of this showing the distribution of the Steitzviridae family of viruses (bottom).

\section{Tokenizer Details}
\label{app:tokenizer-details}

Our tokenizer implementation is adapted from \texttt{minbpe}\footnote{\url{https://github.com/karpathy/minbpe}}. It is trained on a subset of sequences consisting of 2 billion base pairs. These sequences are uniformly sampled from all of the available wastewater sequencing runs from our data sources. Similarly to BPE tokenizers trained on natural language datasets, we treat the beginning of each sequence differently, in our case by prepending a `\_' character to the beginning of each read. During pretraining, we postpend a \texttt{[BOS]} token to separate each sequence. Our tokenizer consists of the following special tokens: \texttt{[PAD]}, \texttt{[UNK]}, \texttt{[SEP]}, \texttt{[BOS]}, \texttt{[EOS]}, and \texttt{[MASK]} to allow for diverse applications during fine-tuning. In total, it has of a vocabulary size of 1024. 

In our preliminary experiments, we also experimented with a larger vocabulary size of 4096, but due to length characteristics of our metagenomic data, this design choice results in many short tokenized sequences that may not be able to provide meaningful learning signal. We thus decided to move forward with a vocabulary size of 1024 to balance efficiency and downstream performance. 
\clearpage
\section{Additional Experimental Details}

\subsection{Additional Details for the Pathogen Detection Benchmark}
\label{app:pathogen-details}

In Table~\ref{tab:pathogen-hyperparameter}, we show our choices of hyperparameters for fine-tuning experiments. 

\begin{table}[h]
    \small  
    \centering
    \begin{tabular}{lccccc}
        \toprule
        DNABERT-$\star$ & \multicolumn{5}{c}{Full Model} \\
        NT-$\star$ & \multicolumn{5}{c}{LoRA} \\
        METAGENE-1 & \multicolumn{5}{c}{LoRA} \\
        \midrule
        LoRA Modules & \multicolumn{5}{c}{query, key, value, dense} \\
        LoRA Rank & \multicolumn{5}{c}{8} \\
        LoRA $\alpha$ & \multicolumn{5}{c}{16} \\
        LoRA Dropout & \multicolumn{5}{c}{0.1} \\
        \midrule
        Optimizer & \multicolumn{5}{c}{AdamW} \\
        Optimizer Momentum & \multicolumn{5}{c}{$\beta_1$, $\beta_2$ = 0.9, 0.999} \\
        Learning Rate & \multicolumn{5}{c}{1e-4$^\Lambda$}\\
        LR Scheduler & \multicolumn{5}{c}{Linear Warmup + Constant LR} \\
        Warmup Steps & \multicolumn{5}{c}{50} \\
        Weight Decay & \multicolumn{5}{c}{0.01} \\
        Denominator $\epsilon$ & \multicolumn{5}{c}{1e-8} \\
        Precision & \multicolumn{5}{c}{BF16-mixed} \\
        \midrule
        Batch Size & \multicolumn{5}{c}{32} \\
        Epochs & \multicolumn{5}{c}{10} \\
        Hardware & \multicolumn{5}{c}{NVIDIA A100 80GB} \\
        \bottomrule
    \end{tabular}
    \vspace{3mm}
\caption{Hyperparameter settings for the Pathogen Detection fine-tuning experiments. $\Lambda$: for DNABERT-S, we halve the learning to 5e-5 as we observe clear oscillation behavior in the training loss.}
    \label{tab:pathogen-hyperparameter}
\end{table}

\clearpage

\subsection{Additional Details for the GUE Benchmark}
\label{app:gue-details}

In Table~\ref{tab:gue-hyperparameter}, we show our choices of hyperparameters for fine-tuning experiments. 

\begin{table}[h]
    \small  
    \centering
    \begin{tabular}{lccccc}
        \toprule
        LoRA Modules & \multicolumn{5}{c}{query, key, value, dense$^\Lambda$} \\
        LoRA Rank & \multicolumn{5}{c}{8} \\
        LoRA $\alpha$ & \multicolumn{5}{c}{16} \\
        LoRA Dropout & \multicolumn{5}{c}{0.1} \\
        \midrule
        Optimizer & \multicolumn{5}{c}{AdamW} \\
        Optimizer Momentum & \multicolumn{5}{c}{$\beta_1$, $\beta_2$ = 0.9, 0.999} \\
        Learning Rate & \multicolumn{5}{c}{\{1e-4 $\cdots$ 1e-3\}$^\Omega$}\\
        LR Scheduler & \multicolumn{5}{c}{Linear Warmup + Constant LR} \\
        Warmup Steps & \multicolumn{5}{c}{50} \\
        Weight Decay & \multicolumn{5}{c}{0.01} \\
        Denominator $\epsilon$ & \multicolumn{5}{c}{1e-8} \\
        Precision & \multicolumn{5}{c}{BF16-mixed} \\
        \midrule
        Batch Size & \multicolumn{5}{c}{32} \\
        Epochs & \multicolumn{5}{c}{10} \\
        Hardware & \multicolumn{5}{c}{NVIDIA A100 80GB} \\
        \bottomrule
    \end{tabular}
    \vspace{3mm}
\caption{Hyperparameter settings for the GUE fine-tuning experiments. $\Lambda$: LoRA is applied to query-value or query-key-value-dense modules. $\Omega$: learning rates are tuned over a equally-spaced grid of 1e-4, 2e-4, $\cdots$, 1e-3. All hyperparameters are selected according to performances on validation sets.}
    \label{tab:gue-hyperparameter}
\end{table}

\appendix

\newpage
\end{document}